\documentclass[11pt]{article}
\def\D{{\mathscr D}}
\def\half{\frac{1}{2}}

\def\slash#1{\, /\kern-0.6em{#1}}

\usepackage{bm}
\input mathrsfs.sty
\begin{document}

\begin{center}
\vskip 0.0cm
\large{\bf Confinement of monopole using flux
string\footnote{Presented by C.Chatterjee at 17th DAE-BRNS High
 Energy Physics Symposium (HEP06), Kharagpur, India, 11-16 Dec
 2006.}}
\vskip 0.0cm {\normalsize Chandrasekhar Chatterjee, Amitabha
Lahiri}\\{\normalsize{S.N.Bose National Centre for
  Basic sciences\\
Block-JD, Sector-III, Salt Lake\\Kolkata-700098\\}}

\end{center}        

\begin{abstract}
We study the confinement of fermionic magnetic monopoles by a
thin flux tube of the Abelian Higgs model. Parity demands that
the monopole currents be axial. This implies that the model is
consistent only if there are at least two species of fermions
being confined.
\end{abstract}

In Nambu's model of confinement \cite{Nambu:1974zg}, quarks are the
endpoints of an open Dirac string~\cite{Dirac:1948um}.  For massive
gauge fields these strings become real \cite{Balachandran:1975qc},
unlike the usual Dirac string. We take fermionic monopoles and
attach them to the ends of a flux string such that no flux can
escape.  To do this, we first dualize the theory to write it in
terms of a `magnetic' gauge field and an antisymmetric tensor. Then
we minimally couple the monopole current to the `magnetic' photon
and redualize the theory.  The result is a theory of magnetic flux
tubes interacting with a massive Abelian vector gauge field. The
tubes are sealed at the ends by fermions, thus providing a toy
model for quark confinement.

We start with the generating functional for the Abelian Higgs model
in $3+1$ dimensions, coupled to an Abelian gauge field $A^e_\mu$.
The partition function is given by $Z = \int \D
A^e_\mu\D\Phi\D\Phi^* \exp iS\,,$ where
\begin{eqnarray}
S
&=& \int d^4 x \left(- \frac14 F^e_{\mu\nu}F^{e\mu\nu} +
|D_\mu\Phi|^2 + \lambda (|\Phi|^2 - v^2)^2 \right)\,,
\label{flux.Higgs}
\end{eqnarray}
where $D_\mu = \partial_\mu +i e A^e_\mu\,,$ and $F^e_{\mu\nu} =
\partial_\mu A^e_\nu - \partial_\nu A^e_\mu$ is the Maxwell field
strength.

We will consider the theory in the London limit $\lambda\rightarrow
\infty\,, |\Phi|= v\,.$ We change variables from $\Phi\,,$
$\Phi^*\,$ to the radial Higgs field $\rho$ and the angular field
$\theta\,.$ Then the action becomes $\int d^4 x \left(- \frac14
F^e_{\mu\nu}F^{e\mu\nu} + \frac{v^2}2 (\partial_\mu\theta + e
A^e_{\mu})^2 \right)\,.$ In the presence of flux tubes we can
decompose $\theta$ into a regular and a singular part. The singular
part $\theta_s$ is related to the world sheet $\Sigma$ of the flux
tube as $\epsilon^{\mu\nu\rho\lambda}\partial_{\rho}
\partial_{\lambda}\theta^s = \Sigma^{\mu\nu}\,,$ where
\begin{eqnarray}
\Sigma^{\mu\nu} &=&
2\pi n\int_{\Sigma}d\sigma^{\mu\nu}(x(\xi))\,\delta^4(x-x(\xi))\,,
\label{flux.sigma}
\end{eqnarray}
with $\xi = (\xi^1, \xi^2)$ the coordinates on the world-sheet,
$d\sigma^{\mu\nu}(x(\xi)) = \epsilon^{ab}\partial_a x^\mu
\partial_b x^\nu\,,$ and $2\pi n $ is the vorticity
\cite{Marino:2006mk}.

We can rewrite the action using standard techniques of
linearization~\cite{Davis:1988rw,Mathur:1991ip,Lee:1993ty}. Then
integrating over $\theta_r$ and $A^e_{\mu}$ we get the partition
function as
\begin{equation}
\int \D A^m_{\mu} \D x_{\mu}(\xi) \D B_{\mu\nu}
\exp i\int\left[-
\frac 14 (e B_{\mu\nu} + \partial_{[\mu}A^m_{\nu]})^2
+ \frac 1{12 v^2} H_{\mu\nu\rho}^2 -
\half \Sigma_{\mu\nu}B^{\mu\nu} \right] \,,
\label{flux.functional}
\end{equation}
where $H_{\mu\nu\rho} = \partial_\mu B_{\nu\rho} + \partial_\nu
B_{\rho\mu} + \partial_\rho B_{\mu\nu}$. What we have achieved here
is the dualization of the photon $A^e_{\mu}$ to the `magnetic
photon' $A^m_{\mu}$.  We have also replaced the integration measure
$\D\theta^s$ by $\D x_\mu(\xi)$.  Here $x_{\mu}(\xi)$ parametrizes
the surface on which the field $\theta$ is singular. The Jacobian
for this change of variables gives the action for the string on the
background spacetime~\cite{Polchinski:1991ax,Akhmedov:1995mw}.

The equation of motion for the field $B_{\mu\nu}$ is
$\partial_\lambda H^{\lambda\mu\nu} = -(m^2/e)\, G^{\mu\nu} - m^2
\,\Sigma^{\mu\nu} \,, $ where $G_{\mu\nu}= e B_{\mu\nu} +
\partial_{\mu}A^m_{\nu} - \partial_{\nu}A^m_{\mu}\,,$ and $m =
ev$. Absence of a magnetic current gives $\partial_\mu G^{\mu\nu} =
0$, so from the above equation of motion we get $\partial_\nu
\Sigma^{\mu\nu} = 0 $. This equation means that in the absence of
magnetic monopoles, the vorticity current tensor $\Sigma_{\mu\nu}$
is conserved. Therefore in the absence of monopoles, flux tubes are
closed or infinite.

Now consider a massless fermionic monopole current minimally
coupled to the magnetic or dual photon. This current is axial in
order to conserve the parity of Maxwell's equations. However, a
theory containing axial currents is anomalous. We can cancel the
anomaly by introducing another species of fermionic monopoles with
opposite charge. Writing the two species as $q$ and $q'\,,$ with
magnetic charges $+g$ and $-g\,,$ we can write the current as
$j^\mu_m = g\bar q\gamma_5 \gamma^\mu q - g\bar q'\gamma_5
\gamma^\mu q' $.

The partition function of Eq.~(\ref{flux.functional}) is modified
to include the fermionic monopoles, minimally coupled to the
`magnetic photon' $A_\mu^m\,,$ so the Lagrangian now reads
${\mathscr L} = - \frac 14 (e B_{\mu\nu} +
\partial_{[\mu}A^m_{\nu]})^2 + ( 1/{12 v^2}) H_{\mu\nu\rho}^2 -
\half \Sigma_{\mu\nu}B^{\mu\nu}+ i \bar q \slash\partial q + i \bar
q' \slash\partial q' - A^m_\mu j^\mu_m \,.$ ~The conservation
condition is modified to $(1/e) \partial_\nu \Sigma^{\mu\nu} =
j^\mu_m $.

We can see that this equation is a consequence of gauge invariance.
We take a transformation $B_{\mu\nu} \rightarrow B_{\mu\nu} +
\partial_{\mu}\Lambda_{\nu} - \partial_{\nu}\Lambda_{\mu} \,,
A^m_{\mu} \rightarrow A^m_{\mu} - (k/g) \Lambda_{\mu} \,. $ The
Lagrangian is invariant if we set $eg = k$. The flux due to the
monopole is $4\pi g$. Since this flux is fully confined inside the
tube, we see by using the conservation condition and $eg=k$ that,
$2 n \pi = \mbox{vorticity flux}\; = (k/g)~(\mbox{monopole flux})\;
= 4\pi k$. So $\displaystyle k = (n/2)$, and we must have the Dirac
quantization condition $\displaystyle eg = (n/2)$. Since the
transformation given above is only a change of variables, $Z$
cannot depend on $\Lambda_\mu$. Thus $\Lambda_{\mu}$ can be
integrated out with no effect other than the introduction of an
irrelevant constant factor in $Z$, which we ignore. After
integrating over $\Lambda_\mu\,,$ we get \cite{chandra}
\begin{eqnarray}
Z = \int \D A^m_\mu\cdots
\delta\Big[\frac 1e \partial_\mu\Sigma^{\mu\nu} +  j^\nu_m\Big]
\exp i\int d^4x \left[- \frac 14 (e B_{\mu\nu} +
\partial_{[\mu}A^m_{\nu]})^2  + \frac 1{12 v^2} H_{\mu\nu\rho}^2
\right.
\nonumber\\
\qquad\qquad \left. - \half\Sigma_{\mu\nu}B^{\mu\nu}
+ i \bar q \slash\partial q + i \bar q'
\slash\partial q' - A^m_\mu j^\mu_m
\right]\,.
\end{eqnarray}
One can see from the $\delta$-functional that the vorticity current
tensor is not conserved, but is cancelled by the fermionic current.
So the strings are open strings with fermions stuck at the ends.
Now we dualize the theory a second time and get back to a vector
gauge field. Then the partition function becomes
\begin{eqnarray}
Z &=& \int \D x_\mu(\xi) \D B_{\mu\nu} \D A_\mu\cdots \delta
\Big[\frac 1e\partial_\mu\Sigma^{\mu\nu} + j^\nu_m\Big] \exp
\,i\int d^4x \Big[- \frac 14 F_{\mu\nu}F^{\mu\nu} \nonumber\\ && +
\qquad \frac 1{12 v^2} H_{\mu\nu\rho} H^{\mu\nu\rho} + \frac 1{2g}
\epsilon^{\mu\nu\rho\lambda} B_{\mu\nu} \partial_\rho A_\lambda + i
\bar q \slash\partial q + i \bar q'\slash\partial q' \Big]\,.
\label{mono.final}
\end{eqnarray}
Here $F_{\mu\nu} = \partial_\mu A_\nu - \partial_\nu A_\mu -
\displaystyle(1/2e) \epsilon_{\mu\nu\sigma\lambda}
\Sigma^{\sigma\lambda}\,.$ The theory is now in the form we
originally intended, and contains thin tubes of flux.  The new
feature is that the ends of the flux tube are {\em sealed} by
fermions, so that no flux escapes.

The mass per unit length of these strings is easily found to be
$\mu \sim g^2 v^2 \lambda$. Such a string of finite length would
collapse in order to minimize the energy unless it was stabilized
by its angular momentum.  For a rotating string of length $l,$
energy per unit length $\mu$, angular momentum $J,$ the energy
function is $E = \mu l + J^2/2\mu l^3$.  We can see that for the
stable flux tube with magnetic monopoles at the ends, ${J}/{E^2} =
constant $, similar to the well-known Regge trajectory for mesons.
The gauge field $A_\mu$ is massive, with mass
$m=v/g$~\cite{Cremmer:1973mg,Allen:1990gb}. It does not couple
directly to the fermionic monopoles at the ends.  The
$\delta$-functional guarantees that the monopoles must seal the
ends of the string. If we suggestively rename $q$ and $q'$ to $u$
and $\bar d\,,$ the allowed configurations are $u\bar d\,, \bar u
d\,,$ and $u\bar u \pm d\bar d\,,$ which can couple to electroweak
gauge fields also.
%%%%%%%%%%%%%%%%%%%%%%%%%
%\acknowledgements{}
%%%%%%%%%%%%%%%%%%%%%%%%%
%It is a pleasure to acknowledege useful discussions with M.~Mathur,
%P.~B.~Pal and R.~Banerjee.

%%%%%%%%%%%%%%%%%%%%%%%%%%%%%%%%%%%%%%%%%%%%%%%%%%%%%%%%%%%%%%%%%%%%

%%%%%%%%%%%%%%%%%%%%%%%%%%%%%%%%%%%%%%%%%%%%%%%%%%

\end{document}